\documentclass[trackchanges,twocolumn]{aastex701}

\hypersetup{citecolor=blue, filecolor=blue, urlcolor=blue}

\usepackage{braket}
\usepackage{amsmath}
\usepackage{bm}
\usepackage{rotating}
\usepackage{comment}
\usepackage{threeparttable}

\newcommand{\lya}{Ly$\alpha\,$}

\newcommand{\heii}{\ion{He}{2}}

\newcommand{\oiii}{[\ion{O}{3}]}

\begin{document}

\title{The Light Echo of a High-Redshift Quasar mapped with Lyman-$\alpha$ Tomography}

\author[0000-0003-2895-6218]{Anna-Christina Eilers}
\affiliation{Department of Physics, Massachusetts Institute of Technology, Cambridge, MA 02139, USA}
\affiliation{MIT Kavli Institute for Astrophysics and Space Research, Massachusetts Institute of Technology, Cambridge, MA 02139, USA}
\email[show]{eilers@mit.edu} 

\author[0000-0002-5367-8021]{Minghao Yue}
\affiliation{MIT Kavli Institute for Astrophysics and Space Research, Massachusetts Institute of Technology, Cambridge, MA 02139, USA}
\email{myue@mit.edu}

\author[0000-0003-2871-127X]{Jorryt Matthee}
\affiliation{Institute of Science and Technology Austria (ISTA), Am Campus 1, 3400 Klosterneuburg, Austria}
\email{jorryt.matthee@ist.ac.at}

\author[0000-0002-7054-4332]{Joseph F.\ Hennawi}
\affiliation{Leiden Observatory, Leiden University, P.O. Box 9513, 2300 RA Leiden, The Netherlands}
\affiliation{Department of Physics, University of California, Santa Barbara, CA 93106-9530, USA}
\email{joe@physics.ucsb.edu}

\author[0000-0003-0821-3644]{Frederick B.\ Davies}
\affiliation{Max-Planck-Institut f\"ur Astronomie, K\"onigstuhl 17, D-69117 Heidelberg, Germany}
\email{davies@mpia.de}

\author[0000-0003-3769-9559]{Robert A.\ Simcoe}
\affiliation{Department of Physics, Massachusetts Institute of Technology, Cambridge, MA 02139, USA}
\affiliation{MIT Kavli Institute for Astrophysics and Space Research, Massachusetts Institute of Technology, Cambridge, MA 02139, USA}
\email{simcoe@space.mit.edu}

\author[0000-0003-1534-5186]{Richard Teague}
\affiliation{Department of Earth, Atmospheric, and Planetary Sciences, Massachusetts Institute of Technology, Cambridge, MA 02139, USA}
\email{rteague@mit.edu}

\author[0000-0002-3120-7173]{Rongmon Bordoloi}
\affiliation{Department of Physics, North Carolina State University, Raleigh, 27695, North Carolina, USA}
\email{rbordol@ncsu.edu}

\author[0000-0003-2680-005X]{Gabriel Brammer}
\affiliation{Niels Bohr Institute, University of Copenhagen, Jagtvej 128, Copenhagen, Denmark}
\email{gabriel.brammer@nbi.ku.dk}

\author[]{Yi Kang}
\affiliation{Department of Physics, University of California, Santa Barbara, CA 93106-9530, USA}
\email{yi_kang@ucsb.edu}

\author[0000-0001-9044-1747]{Daichi Kashino}
\affiliation{National Astronomical Observatory of Japan, 2-21-1 Osawa, Mitaka, Tokyo 181-8588, Japan}
\email{daichi.kashino@nao.ac.jp}

\author[0000-0003-0417-385X]{Ruari Mackenzie}
\affiliation{Department of Physics, ETH Z{\"u}rich, Wolfgang-Pauli-Strasse 27, Z{\"u}rich, 8093, Switzerland}
\email{ruari.mackenzie@epfl.ch}

\author[0000-0003-2895-6218]{Rohan P.\ Naidu}
\affiliation{MIT Kavli Institute for Astrophysics and Space Research, Massachusetts Institute of Technology, Cambridge, MA 02139, USA}
\email{rnaidu@mit.edu}

\author[0000-0002-6417-040X]{Benjam\'in Navarrete}
\affiliation{Institute of Science and Technology Austria (ISTA), Am Campus 1, 3400 Klosterneuburg, Austria}
\email{Benjamin.Navarrete@ist.ac.at}

\submitjournal{\apjl}

\begin{abstract}
Ultra-violet (UV) radiation from accreting black holes ionizes the intergalactic gas around early quasars, carving out highly ionized bubbles in their surroundings. Any changes in a quasar's luminosity are therefore predicted to produce outward-propagating ionization gradients, affecting the Lyman-$\alpha$ (Ly$\alpha$) absorption opacity near the quasar's systemic redshift. This ``proximity effect'' is well-documented in rest-UV quasar spectra but only provides a one-dimensional probe along our line-of-sight. Here we present deep spectroscopic observations with the James Webb Space Telescope (JWST) of galaxies in the background of a super-luminous quasar at $z_{\rm QSO}\approx6.3$, which reveal the quasar's ``light echo'' with \lya tomography in the transverse direction. This transverse proximity effect is detected for the first time towards multiple galaxy sightlines, allowing us to map the extent and geometry of the quasar's ionization cone. We obtain constraints on the orientation and inclination of the cone, as well as an upper limit on the obscured solid angle fraction of $f_{\rm obsc}<91\%$. Additionally, we find a timescale of the quasar's UV radiation of $t_{\rm QSO}=10^{5.6^{+0.1}_{-0.3}}$~years, which is significantly shorter than would be required to build up the central supermassive black hole (SMBH) with conventional growth models, but is consistent with independent measurements of the quasars' duty cycle. Our inferred obscured fraction disfavors a scenario where short quasar lifetimes can be explained exclusively by geometric obscuration, and instead supports the idea that radiatively inefficient accretion or growth in initially heavily enshrouded cocoons plays a pivotal role in early SMBH growth. Our results pave the way for novel studies of quasars' ionizing geometries and radiative histories at early cosmic times. 
\end{abstract}


\section{Introduction}

Supermassive black holes (SMBHs) reside at the heart of every massive galaxy, regulating the evolution of their hosts, injecting energy into the circum- and intergalactic media (IGM), and shaping entire Galactic ecosystems over cosmic time. Yet the formation and growth of these SMBHs, in particular at early cosmic times, is still highly debated. At redshift $z\gtrsim 6$, the timescale required to grow a billion solar mass black hole is comparable to the age of the universe when assuming a canonical value for the radiative efficiency of $\epsilon\approx10\%$ for the standard thin accretion disc theory in general relativity \citep{ShakuraSunyaev1973, TanakaHaiman2009, Inayoshi2019}, which has led many studies to invoke massive initial black hole seeds in excess of stellar remnants \citep[e.g.]{Begelman2006}, or suggest rapid, radiatively inefficient accretion episodes \citep[e.g.][]{Inayoshi2016, BegelmanVolonteri2017}. 

Since the luminosity of quasars is powered by accretion onto their central SMBHs \citep{Soltan1982, YuTremaine2002}, we can chronicle the black hole growth by studying the radiative histories of quasars. The quasars' ionizing radiation alters the ionization state of the surrounding intergalactic gas, resulting in a decrease in the opacity to \lya photons, which is known as the ``proximity effect''. This effect is well-documented in rest-UV quasar spectra along our line-of-sight \citep{Fan2006, Carilli2010, Eilers2017a, Eilers2020, Morey2021, Satyavolu2023_rp}, but should also be observable in the transverse direction observed in absorption spectra of background sources intersecting the environment of an ionizing source at close projected distances \citep{Adelberger2004, Schmidt2019}. Due to the finite speed of light $c$, the transverse direction of the proximity effect is sensitive to the quasar’s radiative history, resulting in ``light echos'' that encode the growth history of the SMBH. This implies that the map of the ionization state of the gas at different distances $d$ from the SMBH reflects the ionizing luminosity output of the quasar at a time $t=d/c$. By observing sources located in the background of a foreground quasar, one can study the expected change in the \lya forest opacity at close projected distances to the quasar to tomographically map the quasar's ionization field, which is a technique known as \lya tomography \citep{Lee2014, Schmidt2019, Kakiichi2022}. 

Since galaxies in the background of luminous foreground quasars are often extremely faint, studies of the quasars' transverse proximity effect have so far focused on analyzing quasar pairs, where the environment of a foreground quasar is pierced by a single sightline to a background quasar at close distances projected in the plane of the sky. However, to date only one spectroscopic detection of the transverse proximity effect has been reported using a pair of quasars at $z\sim3$ in the \heii\ \lya forest observed in the spectrum of a background quasar at $z\approx3$ \citep{Jakobsen2003}, while studies of several tens of other quasar pairs did not show the expected opacity change \citep{Schmidt2018}. However, in these studies only a single sightline probes the environment of the foreground quasar, which makes it impossible to disentangle whether short quasar lifetimes or geometric obscuration effects that might prevent the sightline of the background object to be illuminated by the foreground quasar, could be responsible for these non-detections. 

At $z\gtrsim6$ quasars are significantly less abundant than at lower redshifts \citep[e.g.][]{Wang2019, Schindler2023} rendering studies of quasar pairs infeasible. The first detection of a photometric transverse proximity effect along a foreground quasar pierced by a background galaxy sightline has been reported by \citet{Bosman2019} a few years ago, where a narrow-band filter centered around the quasar's redshift at $z_{\rm QSO}\approx5.8$ reveals a putative flux detection along the background sightline. 

Here, we use a sample of twelve galaxies in the background of a luminous quasar in the high-redshift universe to probe the transverse proximity effect for the first time spectroscopically and along multiple sightlines, and create the first three-dimensional map of a quasar's ``light echo'' by means of \lya tomography. We leverage deep spectroscopic observations with the NIRSpec instrument on the James Webb Space Telescope (JWST) of a large number of faint galaxies that were detected behind a luminous quasar along our line-of-sight. 

This paper is organized as follows: In \S~\ref{sec:data} we describe the observations of the galaxy spectra, while \S~\ref{sec:trans_rp} shows the detection of the transverse proximity effect along multiple galaxy sightlines. We set up a model to describe the geometry of the quasar's ionized region and present the first map of the ionized region around the quasar using \lya tomography in \S~\ref{sec:model}. In \S~\ref{sec:discussion} we discuss the implications of our results on the growth of SMBHs in the early universe. 

\begin{figure*}[ht!]
\centering
\includegraphics[width=.49\textwidth]{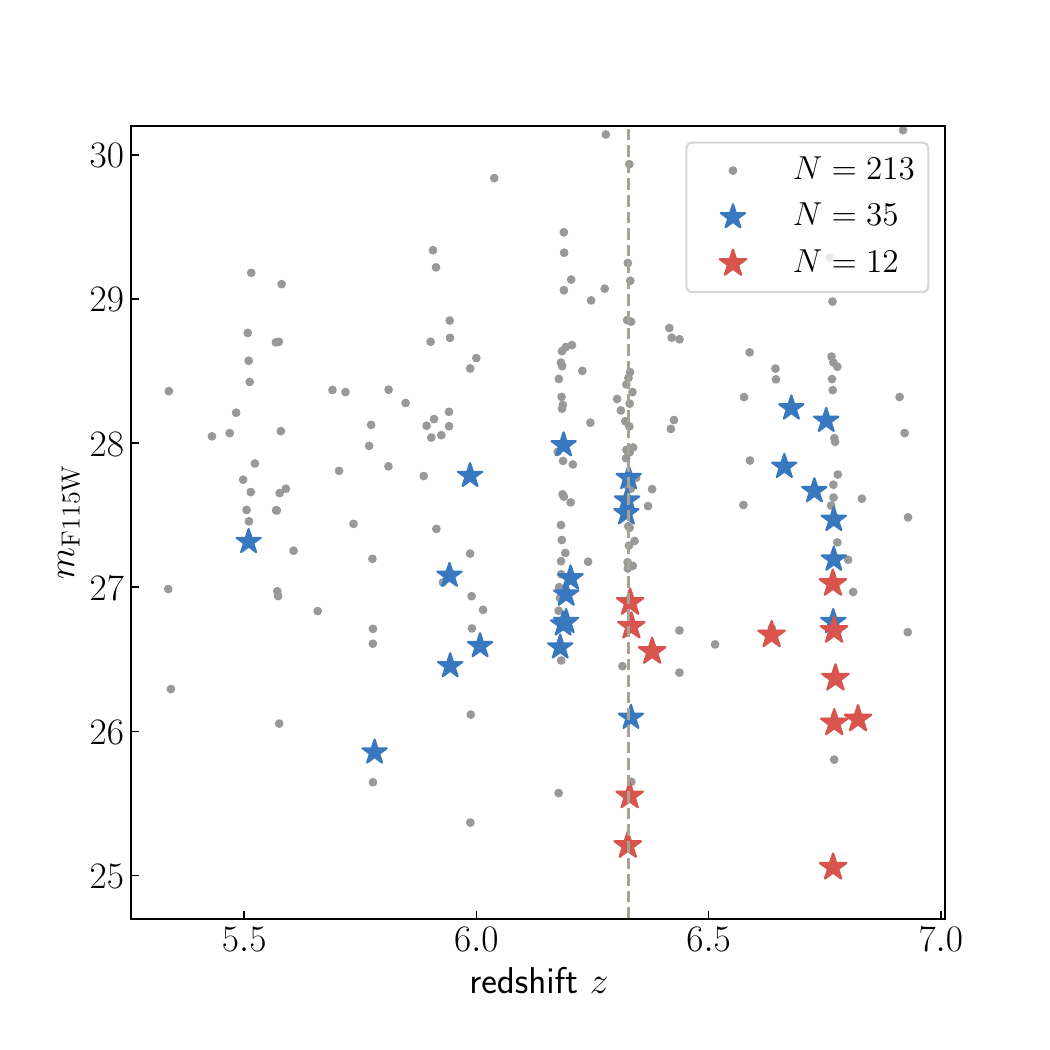}
\includegraphics[width=.49\textwidth]{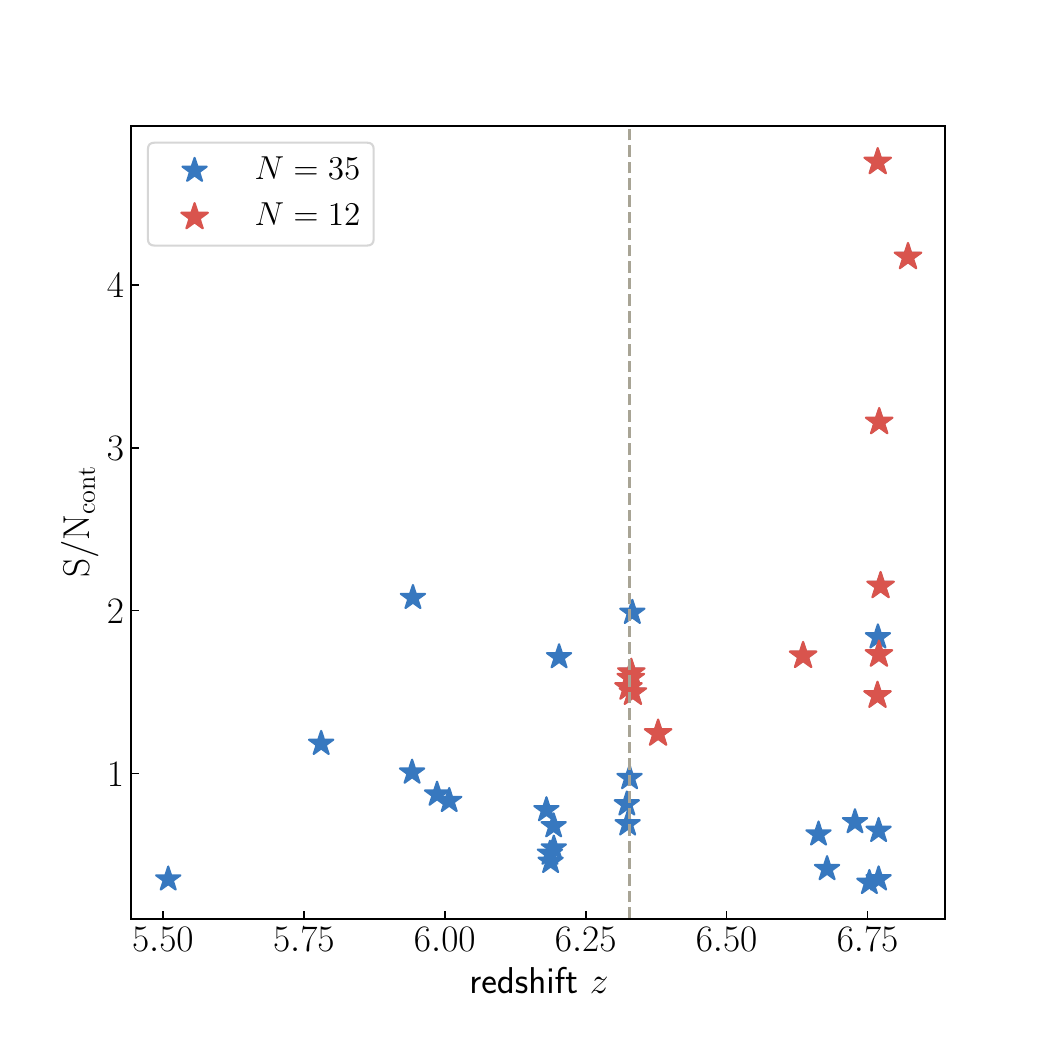}
\caption{
\textbf{Sample of \oiii-emitting galaxies in the quasar field J0100+2802.} A total of $213$ \oiii-emitters (grey points) were previously discovered via NIRCam WFSS observations \citep{Kashino2023} in this quasar field. For $35$ of them (blue stars) we obtained deep NIRSpec/MSA spectroscopic observations, and $12$ of those (red stars) were used in this analysis. Left and right panels show the observed magnitude in NIRCam's F115W filter and the galaxies' continuum signal-to-noise ratio per pixel as a function of redshift $z$, respectively. 
}
\label{fig:O3}
\end{figure*}

\section{Spectroscopic observations of galaxies located behind a luminous quasar}\label{sec:data}

\subsection{Observations \& Data Reduction}

Previously, deep observations with JWST in NIRCam wide-field slitless (WFSS) mode obtained by the EIGER collaboration \citep{Kashino2023} revealed a spectacular overdensity of bright, star-forming galaxies in the background of the super-luminous, high-redshift quasar, J0100+2802 at $z_{\rm QSO}=6.3270$ hosting a $\sim10^{10}\,M_\odot$ SMBHs at its core \citep{Wu2015, Eilers2023, Yue2023}. We obtained follow-up spectroscopic observations using Multi-Object Spectroscopy with the NIRSpec instrument on JWST as part of the \textbf{MASQUERADE} (``Mapping a Super-luminous Quasar's Extended Radiative Emission'') program (program ID: \#4713). Two masks were designed in this quasar field to obtain spectra with the F070LP filter and G140M grating for a subset of $35$ of the in total $213$ detected \oiii-emitters in the quasar field at $5.3\lesssim z_{\rm gal}\lesssim 6.9$, shown in Fig.~\ref{fig:O3}. For the mask design we prioritized bright (${\rm F115W}\lesssim27$~mag) galaxies that lie in the background or environment of the quasar, i.e.\ at $z_{\rm gal}\gtrsim z_{\rm QSO}$. The G140M/F070LP grating/filter combination covers the wavelength range of $0.70\mu{\rm m} <\lambda_{\rm obs}<1.27\mu$m, including the \lya emission at the systemic redshift of the foreground quasar at $\lambda_{\rm obs}\approx8907$~{\AA}, with a nominal resolving power of $R\sim 1000$. The observations were taken in December 2024 with a three slitlet dither pattern in a $7.7$~hours exposure for each mask. Six of these \oiii-emitters were covered by both masks and thus observed for a total of $15.4$~hours. 

The NIRSpec/MSA data are reduced using the \texttt{msaexp} \citep{msaexp} pipeline version 3, the details of which are described in \citep{deGraaff2024, Heintz2025}. In short, the pipeline first requires bias removal and cleaning of the MSA images from ``snowballs'', groups of pixels contaminated by cosmic rays. 
Next, it runs parts of the Level 2 JWST calibration pipeline, including identification of slits, extraction, flatfielding, and photometric calibration. Before combining, each exposure is corrected for wavelength-dependent bar-shadowing using a default $0.35$ minimum value for inverse bar-shadow correction. The background subtraction is done directly on the 2D slit cutouts and estimated in the area outside a box of $9$ pixels centered on the line trace of the source. Finally, we perform an aperture extraction of $13$ cross-dispersion pixels in the image cutouts to obtain the 1D spectrum from combined and rectified exposures. 

\begin{figure*}[ht!]
\centering
\includegraphics[width=0.95\textwidth]{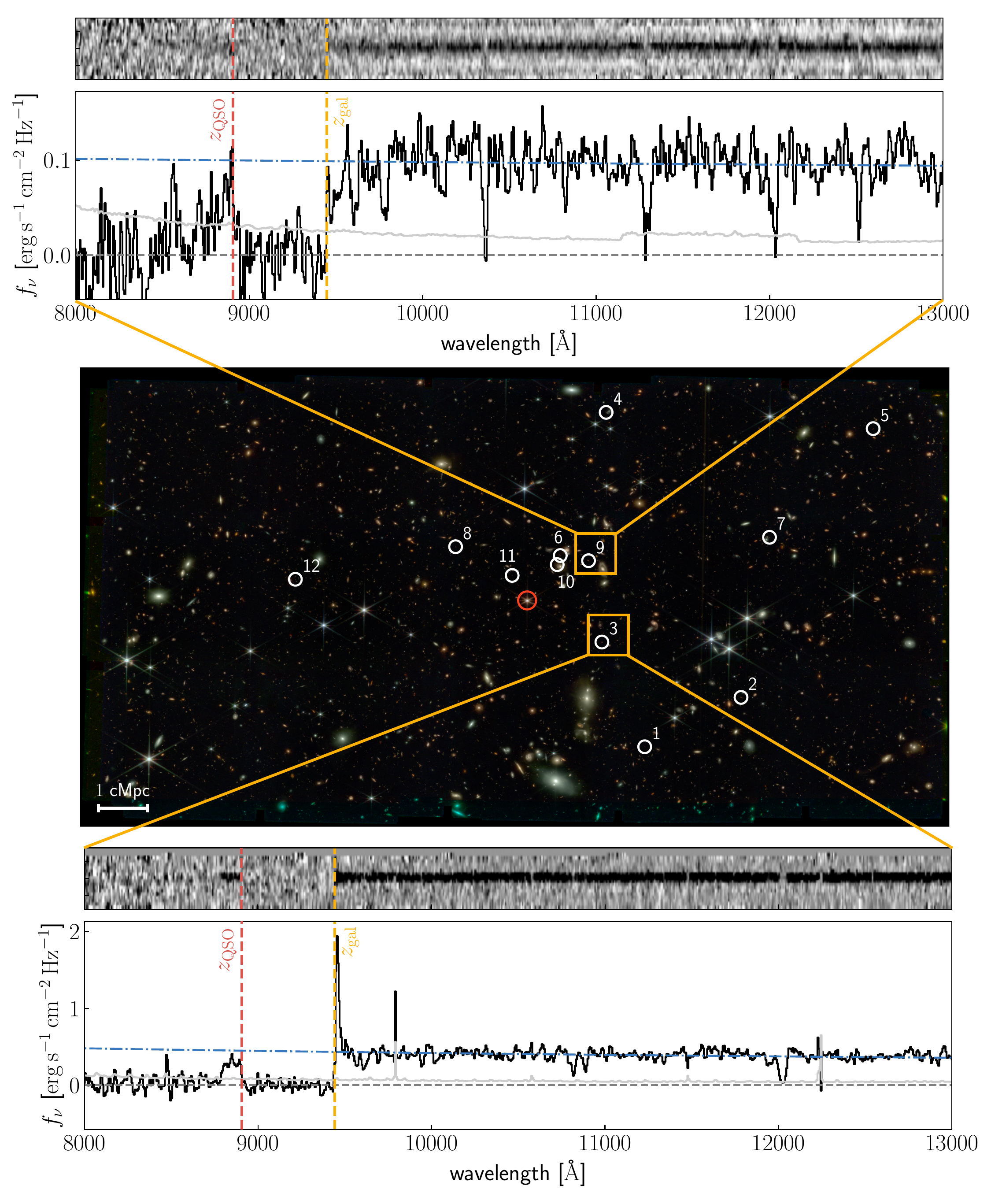}
\caption{\textbf{Observations.} The middle panel shows a color image of the quasar field with the central quasar circled in red, constructed via NIRCam images in the filters F115W, F200W and F356W, showing the spatial locations of the twelve analyzed background galaxies, circled in white. The top and bottom panels show the 2D and 1D spectra of two galaxies, which show a clear detection of the transverse proximity effect at the systemic redshift of the quasar $z_{\rm QSO}$ indicated by the red dashed line. The yellow dashed line shows the galaxies' redshift as determined from the \oiii-emission line, $z_{\rm gal}$, while the blue dash-dotted line demonstrates the power-law fit to the unabsorbed galaxy continuum at $\lambda_{\rm rest} > \lambda_{\rm Ly\alpha}$, which is used to calculate the flux transmission and the optical depth within the \lya forest.}
\label{fig:data}
\end{figure*}

\begin{figure*}[ht!]
\centering
\includegraphics[width=0.49\textwidth]{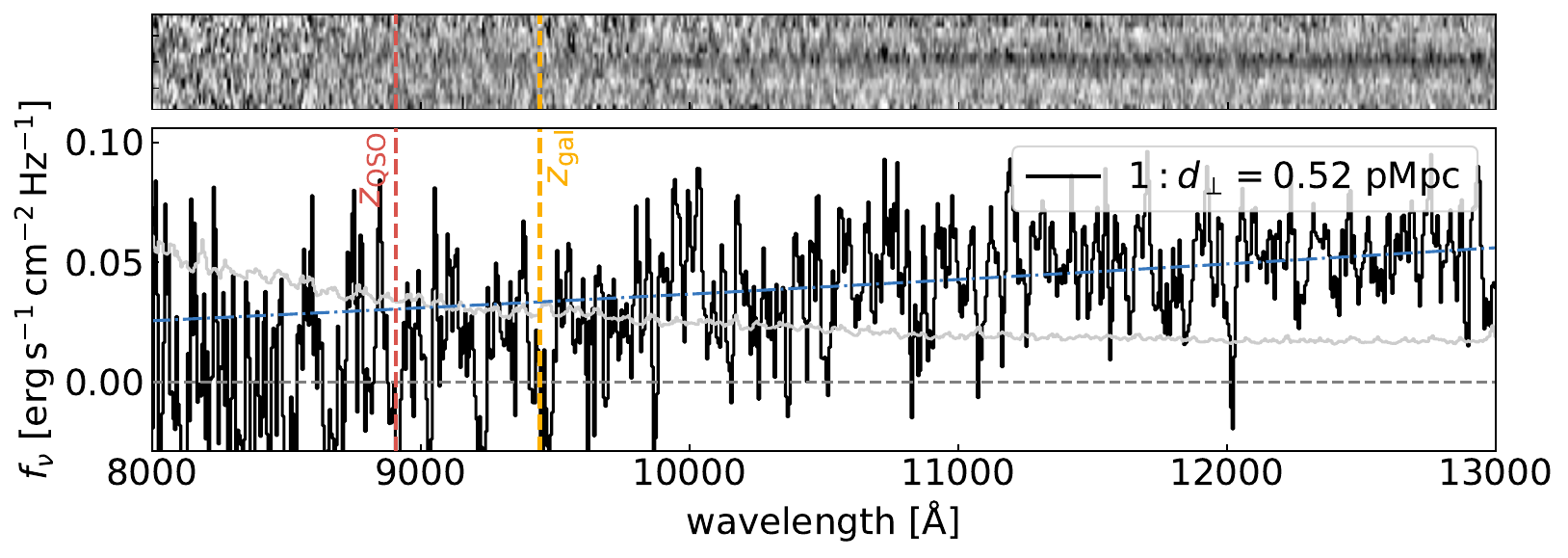}
\includegraphics[width=0.49\textwidth]{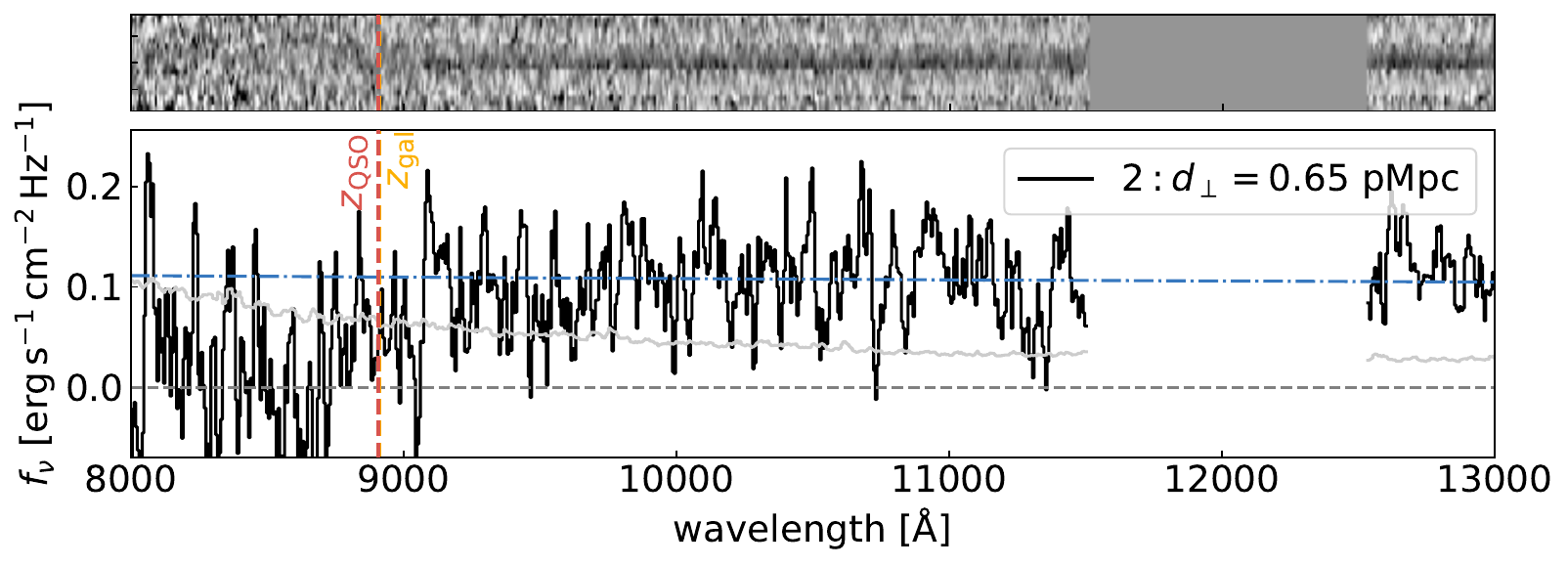}
\includegraphics[width=0.49\textwidth]{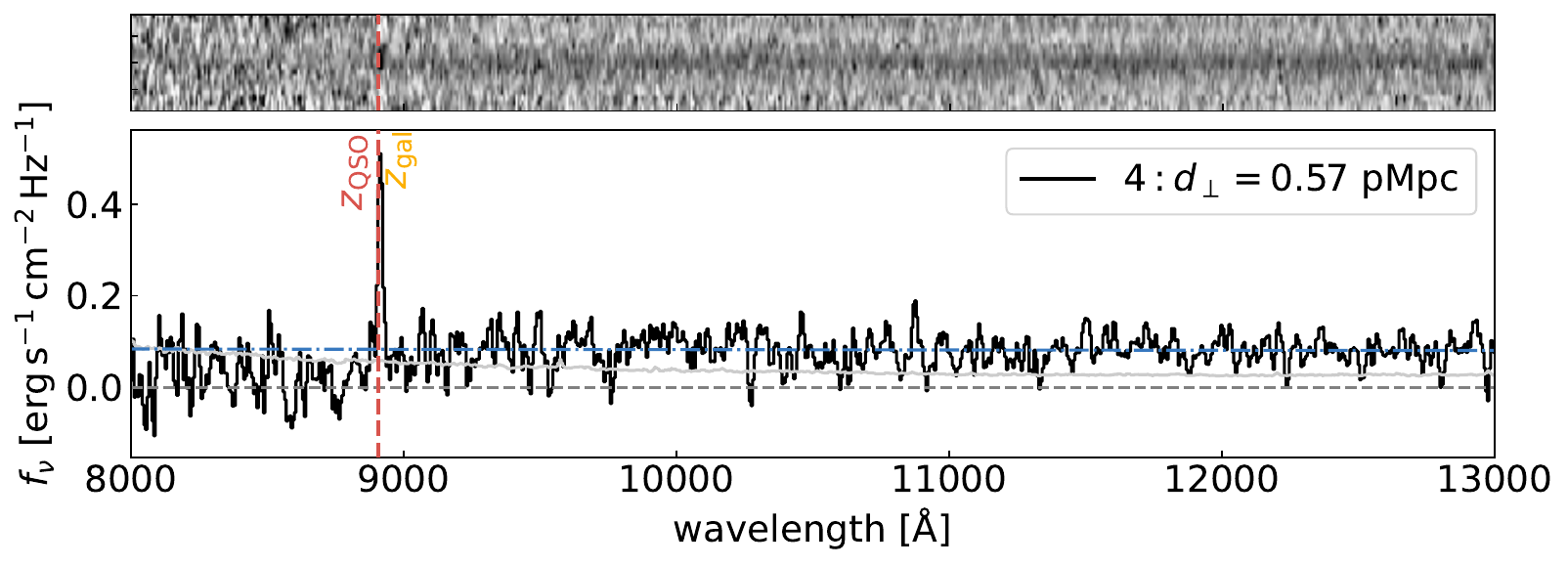}
\includegraphics[width=0.49\textwidth]{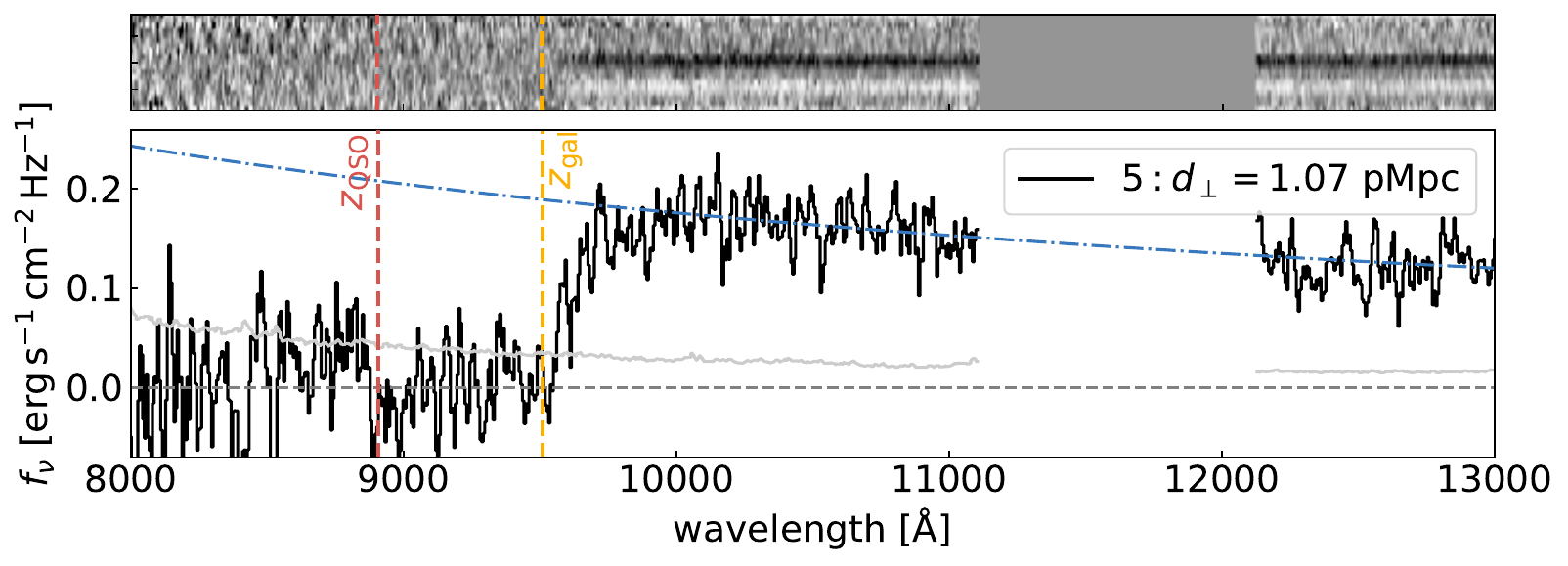}
\includegraphics[width=0.49\textwidth]{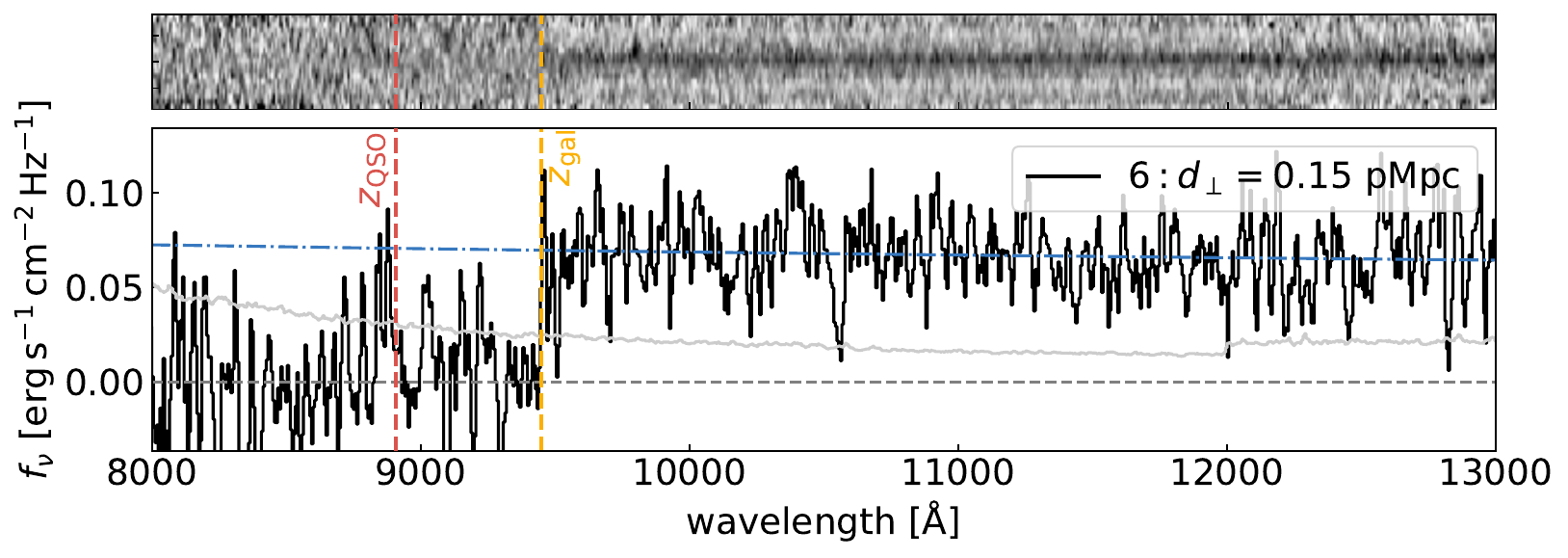}
\includegraphics[width=0.49\textwidth]{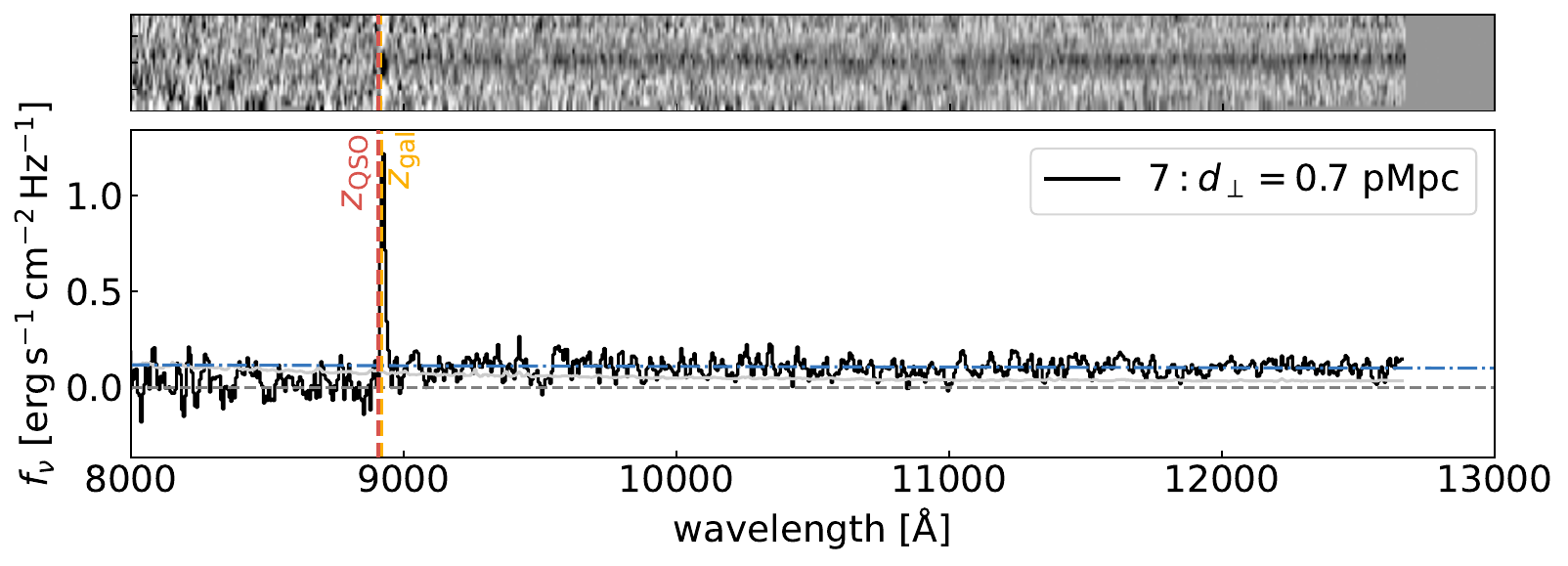}
\includegraphics[width=0.49\textwidth]{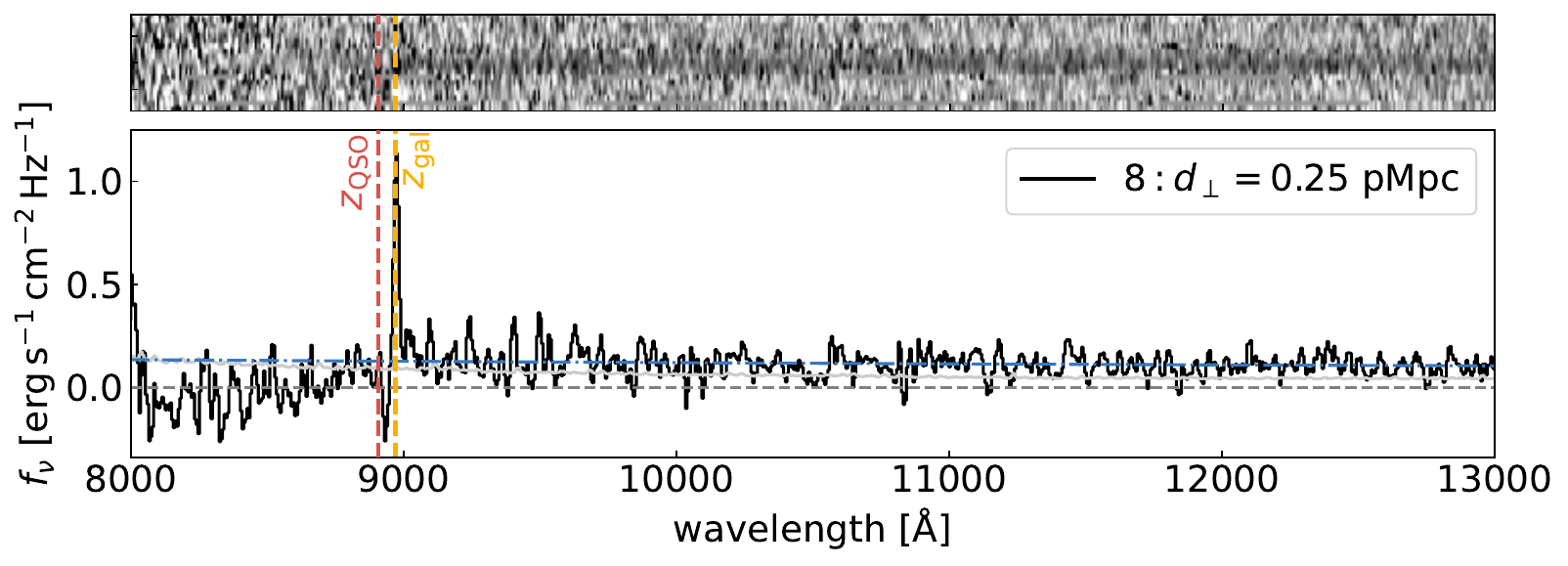}
\includegraphics[width=0.49\textwidth]{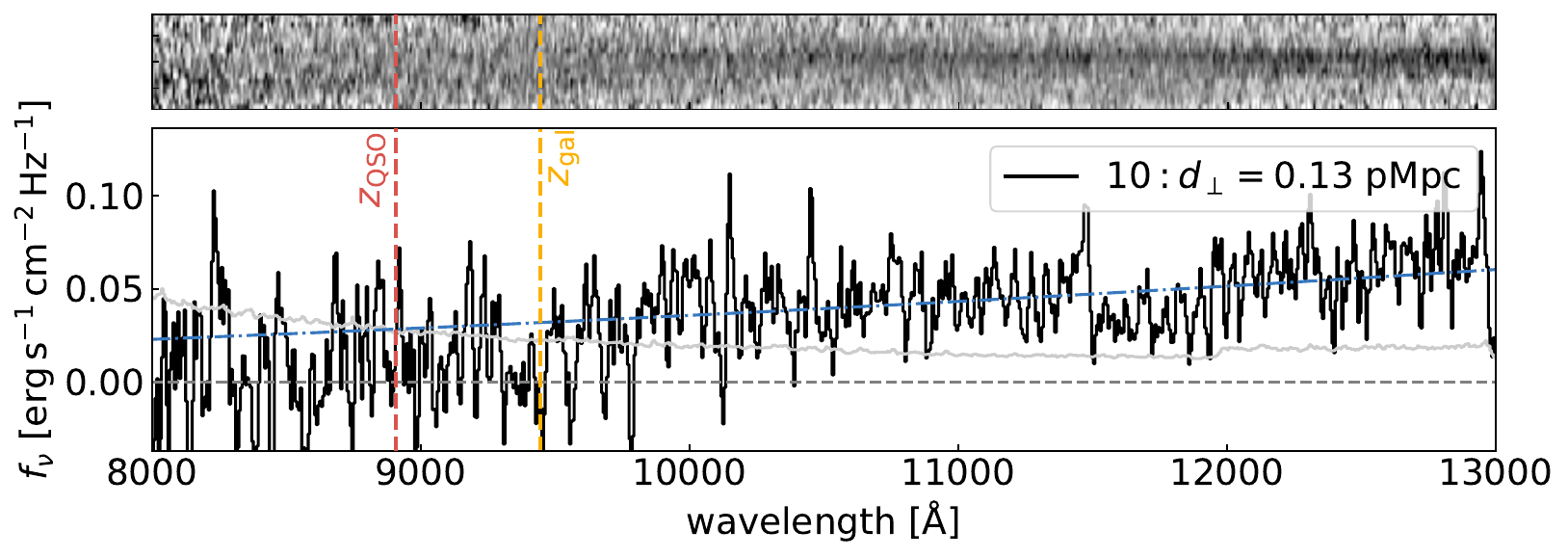}
\includegraphics[width=0.49\textwidth]{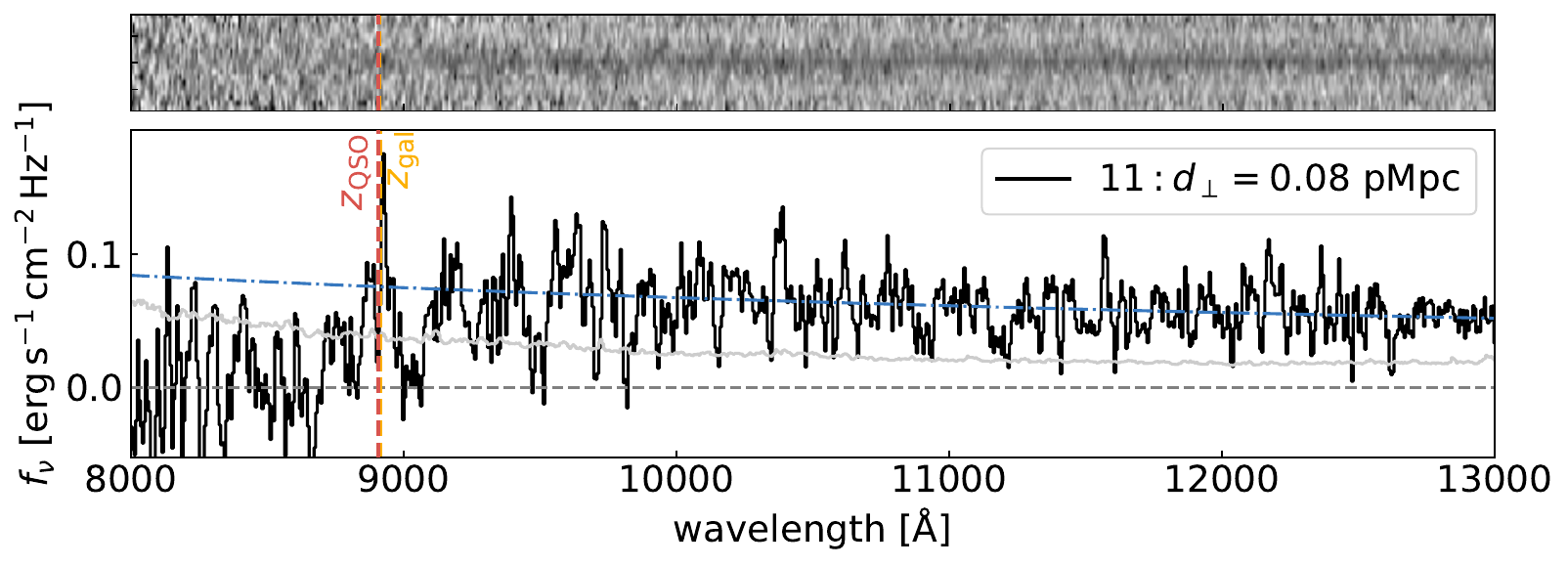}
\includegraphics[width=0.49\textwidth]{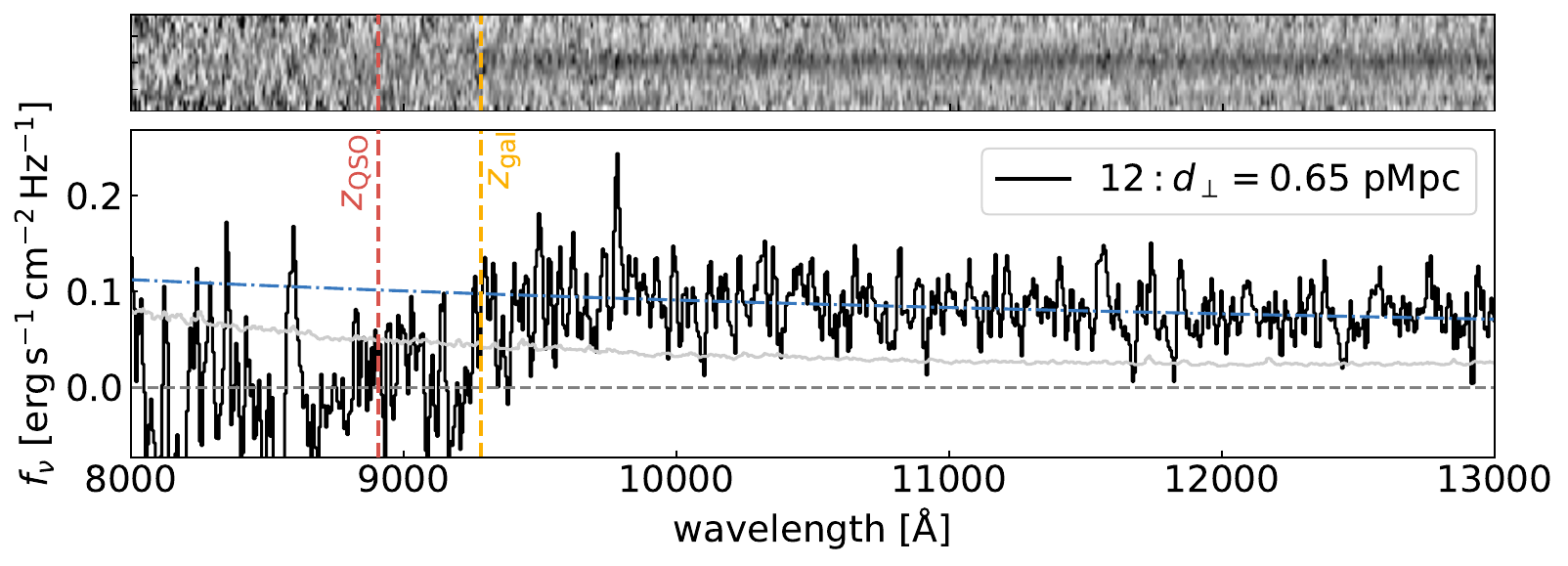}
\caption{
\textbf{Galaxy spectra.} 2D (top) and 1D (bottom) spectra of ten out of the twelve analyzed galaxies in the quasar field (the remaining two galaxy spectra are shown in Fig.~\ref{fig:data}). The red and yellow dashed line indicate the systemic redshift of the quasar $z_{\rm QSO}$ as well as the galaxy's redshift $z_{\rm gal}$, respectively. Blue dashed-dotted lines show the power-law fit to the unabsorbed galaxy continuum emission. 
}
\label{fig:extended_data}
\end{figure*}

\begin{figure*}[ht!]
\centering
\includegraphics[width=0.9\textwidth]{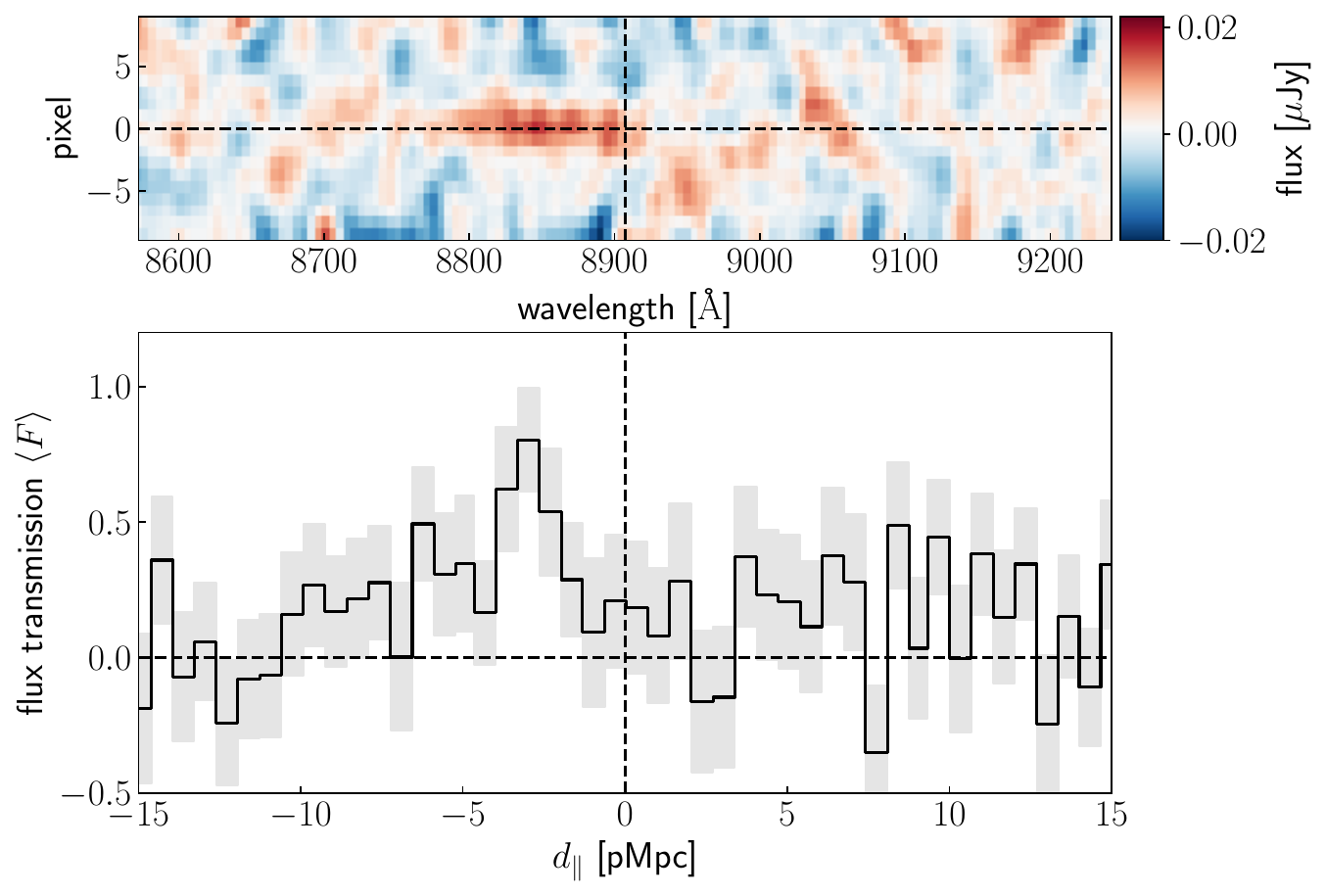}
\caption{
\textbf{Stacked Galaxy Spectra.} The top panel shows the two-dimensional unweighted mean stack of all twelve galaxy spectra colored by the flux. The bottom panel shows the unweighted stack over the same wavelength range of the continuum normalized one-dimensional galaxy spectra binned to $500~\rm km\,s^{-1}$ pixels around the quasar's systemic redshift, which is indicated by the vertical black dashed lines at $d_\parallel=0$~pMpc or $\lambda_{\rm Ly\alpha}\approx 8907$~{\AA}. The horizontal dashed line indicates the spectral trace in the top panel, while it denotes the zero flux level in the bottom panel. Note that the 2D spectra are not normalized by the galaxies' continuum emission before stacking, while we did normalize the 1D spectra to show the stacked \lya transmitted flux. }
\label{fig:stack}
\end{figure*}

\begin{figure*}[ht!]
\centering
\includegraphics[width=.9\textwidth]{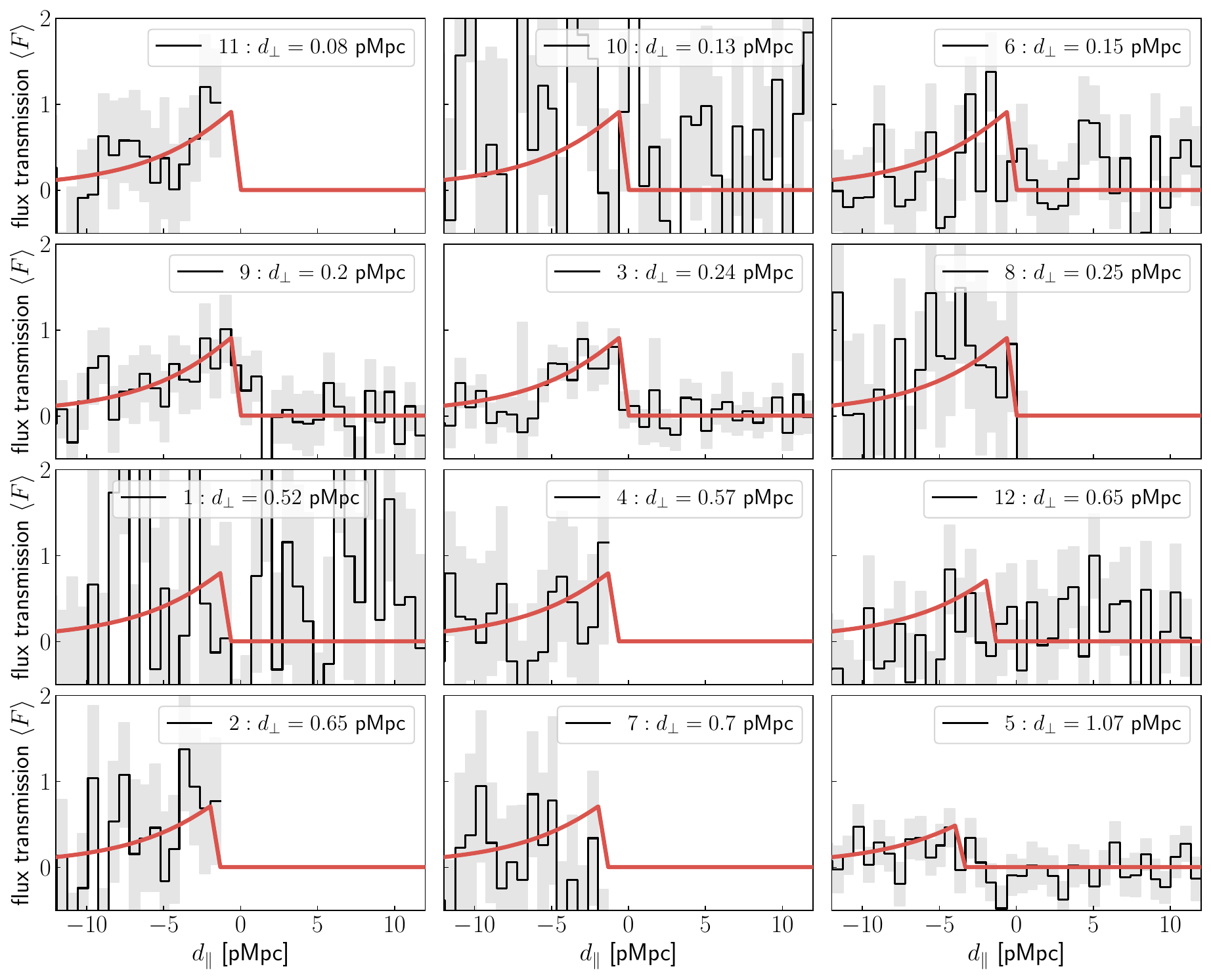}
\caption{\textbf{Flux transmission along individual galaxy sightlines.} The panels show the continuum-normalized, transmitted flux (black) around the systemic redshift of the quasar at $d_\parallel=0$~pMpc with spectral uncertainties (grey) along the $12$ background galaxy sightlines, which are used to model the quasar's ionization cone. The red lines show the expected transmission profile from the best fit of the quasar's ionization cone extracted at each galaxy's location on the sky (see \S~\ref{sec:model}). 
}
\label{fig:bestfit2}
\end{figure*}

\begin{figure*}[ht!]
\centering
\includegraphics[width=\textwidth]{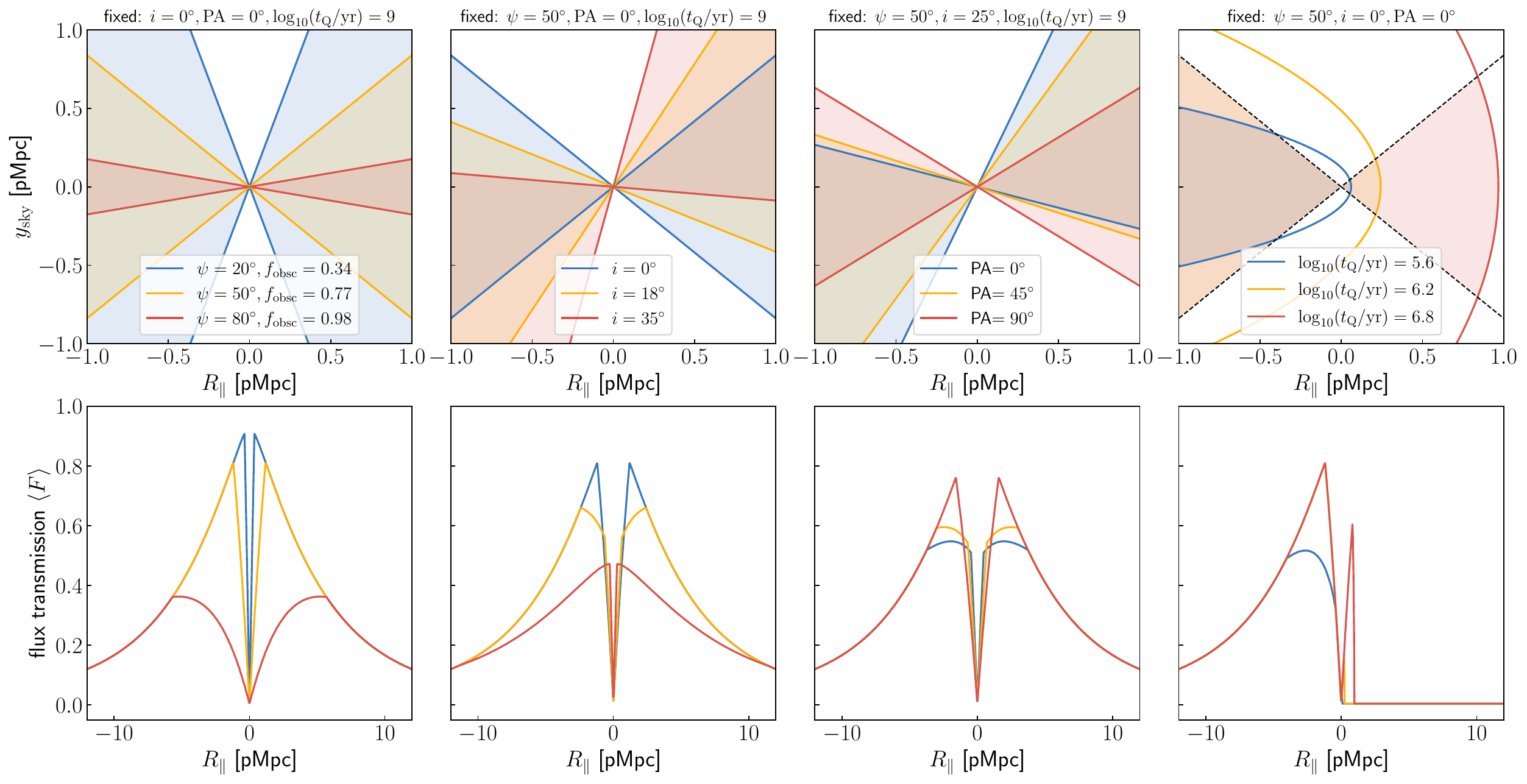}
\caption{
\textbf{Biconical model for quasar's ionization cone.} \textit{Left to right:} panels show the model for the ionization cone with varying parameters, i.e.\ the opening angle of the obscuring medium $\psi$ which corresponds to the solid angle obscured fraction $f_{\rm obsc}$, the inclination of the cone $i$, the position angle $\rm PA$, and the timescale of the quasar's UV-luminous radiation $t_{\rm QSO}$. All other parameters are kept fixed to the values as indicated above each panel. The top panels show the illuminated area around the central quasar, while the bottom panels illustrate the expected mean transmitted flux profile for a highly idealized sample of $1000$ galaxies, which are all aligned and equally distributed along one axis of the sky, $y_{\rm sky}$, while $x_{\rm sky}=0$. 
}
\label{fig:models}
\end{figure*}

\begin{figure}[ht!]
\centering
\includegraphics[width=.45\textwidth]{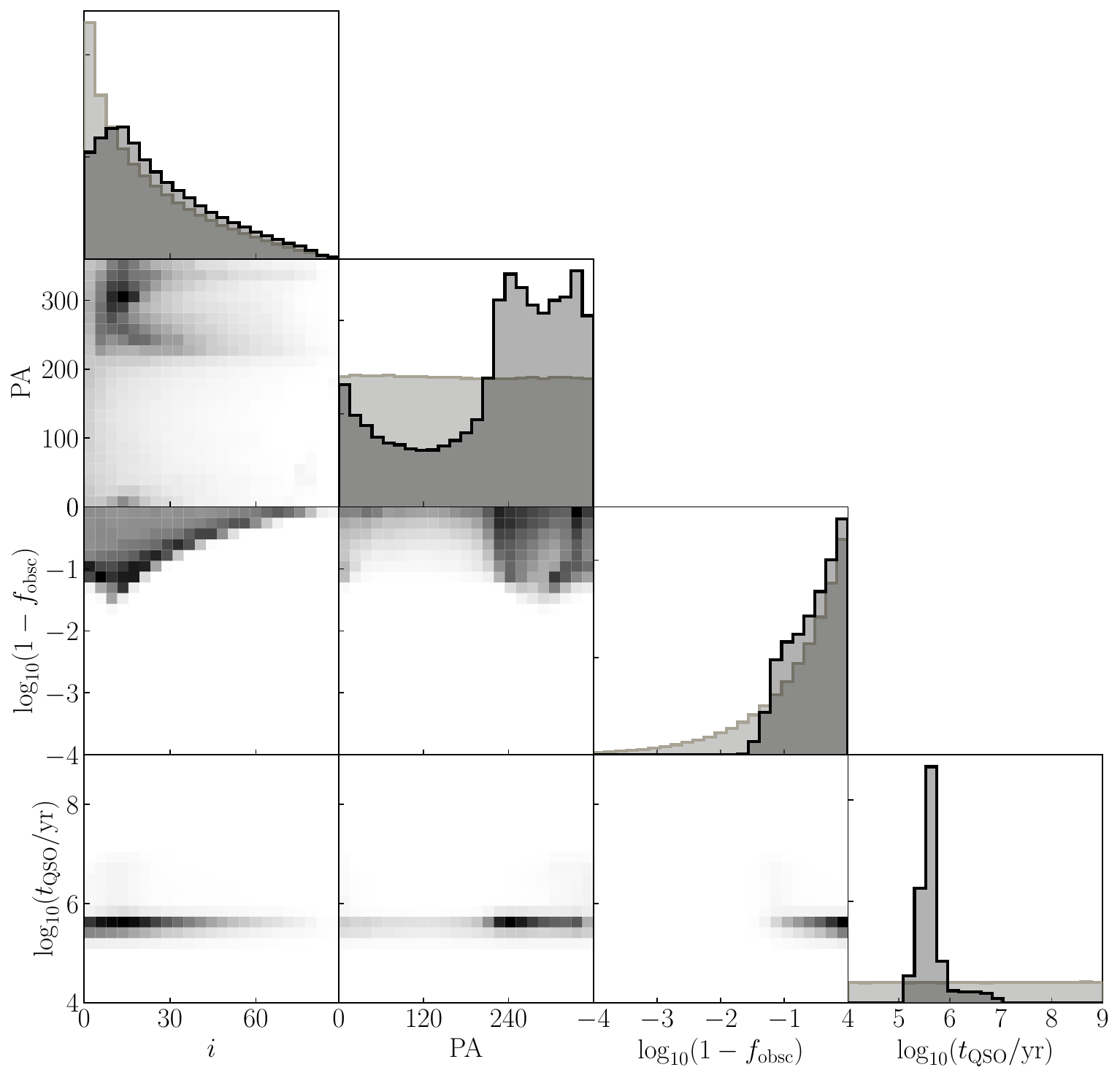}
\caption{
\textbf{Posterior distributions of the model parameters for the ionization cone.} Lower panels shows the marginalized 2D posterior probability distributions determined via MCMC of the four model parameters for the ionization cone, i.e.\ the inclination $i$, the position angle ${\rm PA}$, the unobscured solid angle fraction $\log_{10}(1-f_{\rm obsc})$, and the UV-luminous lifetime of the quasar $\log_{10}(t_{\rm QSO}/\rm yr)$. Light and dark grey histograms in the top diagonal panels show the 1D prior and posterior distributions, respectively. }
\label{fig:corner}
\end{figure}

\begin{figure*}[ht!]
\centering
\includegraphics[width=\textwidth]{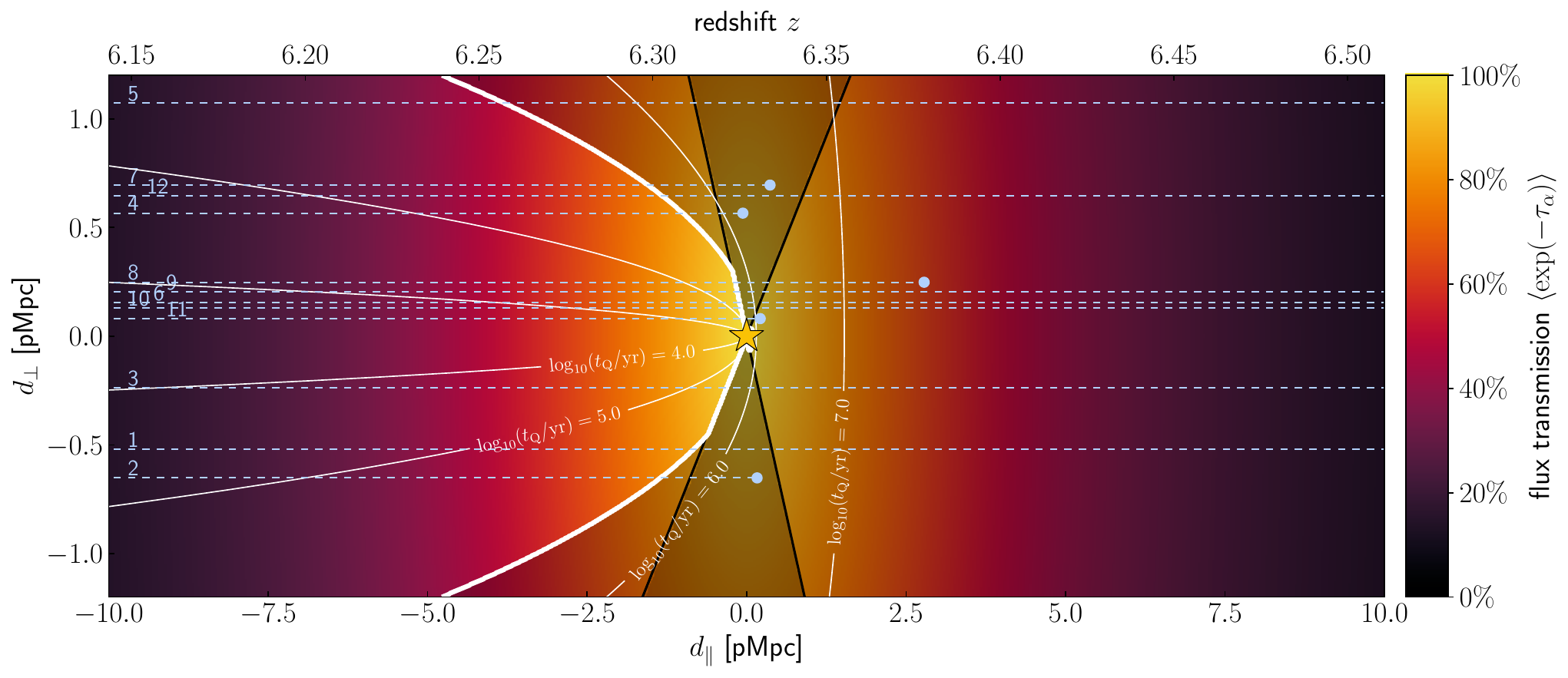}
\caption{
\textbf{Map of the quasar's ionized bubble.} The central quasar is indicated by the yellow star, and all galaxy sightlines shown as blue dashed lines. The underlying colormap indicates the level of the expected \lya flux transmission, while the overplotted black contours illustrate the best fit ionization cone. The thin white parabolas show isochrones, i.e.\ outlines of the quasar's light echo at different times, while the thick white line indicates the best fit to our data of $\log_{10}(t_{\rm QSO}/{\rm yr})=5.7^{+0.1}_{-0.3}$. Thus, all shaded regions are not illuminated by the quasar and only the congruence of the ionization cone and illuminated parabola due to the quasar's light echo is illuminated. 
}
\label{fig:bestfit}
\end{figure*}

\subsection{Flux Continuum Normalization}

The observed galaxies have low metallicities \citep{Matthee2023} and thus we fit a power-law, i.e.\ $f_\lambda\propto(\lambda_{\rm rest}/1500~{\rm{\AA}})^{\beta}$, to the unabsorbed wavelengths between $1300{\rm {\AA}} <\lambda_{\rm rest}<1700$~{\AA} to estimate the galaxies' unabsorbed continuum flux. We extrapolate the power-law to the bluer wavelength region, in order to continuum normalize the flux in the galaxies' \lya forest and estimate the level of flux transmission. At wavelengths close to the galaxies' systemic redshifts any transmitted flux might arise from ionized gas in the IGM or within the galaxies' circumgalactic medium (CGM), while at larger distances any flux transmission arises from highly ionized intergalactic gas. Thus, for galaxies with a redshift close to the systemic redshift of the quasar, we mask conservatively all spectral regions within $\Delta v \leq 1000~\rm km\,s^{-1}$ of the galaxies' \lya emission line, in order to avoid contamination of the flux transmission signal due to ionized gas in the CGM of the galaxies themselves. 

In order to calibrate the flux in the galaxy spectra and account for any potential wavelength dependencies of the NIRSpec data, we compare the NIRSpec/MSA spectrum of the central quasar, J0100+2802, with a well-calibrated ground-based quasar spectrum observed with Folded Port Infrated Echellette (FIRE) \citep{Simcoe2008} spectrograph on the Magellan Telescopes and the X-Shooter spectrograph \citep{Vernet2011} on the Very Large Telescope (VLT). The observations and data reduction of the ground-based quasar spectrum were previously presented \citep{Eilers2023}. We re-scale the NIRSpec/MSA spectrum to match the slope of the ground-based quasar spectrum at wavelengths longer than the \lya emission line and fit a power-law to the ratio of both spectra, which we then use to re-scale all galaxy spectra. The correction factor is small, i.e. $\times1-1.5$, and varies smoothly across the relevant wavelength range, thus resulting in only minor corrections of the galaxy spectra. 

We estimate the signal-to-noise ratio of the galaxies' unabsorbed continuum emission $\rm SN_{\rm cont}$ per pixel between $1300{\rm {\AA}} <\lambda_{\rm rest}<1700$~{\AA}, i.e.\ at wavelengths redwards of their \lya emission line. The detection of the galaxies' continuum emission is essential, in order to search for the expected transmitted \lya flux in the galaxy spectra due to change in IGM opacity. For fourteen galaxies at $z_{\rm gal}\gtrsim z_{\rm QSO}$ we were able to detect the continuum emission with ${\rm SN_{\rm cont}}\geq1$ per $6$~{\AA} pixel, which corresponds to approximately $150-180~\rm km\,s^{-1}$ over the considered wavelength range. However, for two galaxies the \lya emission at the quasar's systemic redshift falls into the chip gap of the NIRSpec detectors, which can therefore not be used for our analysis. 

The resulting twelve objects that meet our selection criteria for this analysis are shown in Fig.~\ref{fig:data} and Fig.~\ref{fig:extended_data}, and their properties are listed in Table~\ref{tab:galaxies}.

\begin{center}
\begin{table*}[h!]
\renewcommand{\arraystretch}{1.2}
\caption{\textbf{Observed galaxies}}
\begin{tabular}{llccccccc} 
galaxy &ID &  RA & DEC& $z_{\rm gal}$ & F115W & $d_\perp$ & ${\rm S/N}_{\rm cont}$ & $\beta$ \\
  && hh:mm:ss.sss & dd:mm:ss.sss & & mag & pMpc & & \\
\hline 
1 & 2084 & 01:00:19.932 & +28:02:16.775 & 6.77 & 27.0 & 0.52 & 1.5 & $-0.38\pm0.44$\\
2 & 4222 & 01:00:20.351 &+28:01:24.026 & 6.33 & 25.6 & 0.65 & 1.6 & $-2.12\pm0.71$\\
3 & 6733 & 01:00:15.822 &+28:02:06.311 & 6.77 & 25.1 & 0.24 & 4.8 & $-2.61\pm0.21$\\
4 & 10998 & 01:00:08.792& +28:01:03.038 & 6.33 & 25.2 & 0.57 & 1.5 & $-2.08\pm0.44$\\
5 & 11981 & 01:00:14.698 &+27:59:17.849 & 6.82 & 26.1 & 1.07 & 4.2 & $-3.45\pm0.18$\\
6 &12217 & 01:00:12.299 &+28:02:00.209 & 6.77 & 26.4 & 0.15 & 2.2 & $-2.24\pm0.29$\\
7 &16479 & 01:00:15.967 &+28:00:29.464 & 6.33 & 26.7 & 0.70 & 1.5 & $-2.33\pm0.45$\\
8 &16759 & 01:00:09.908 &+28:02:40.789 & 6.38 & 26.6 & 0.25 & 1.2 & $-2.52\pm0.54$\\
9 &17385 & 01:00:13.032 &+28:01:49.991 & 6.77 & 26.1 & 0.20 & 3.2 & $-2.15\pm0.22$\\
10 &17577 & 01:00:12.520 &+28:02:03.904 & 6.77 & 26.7 & 0.13 & 1.7 & $0.0\pm0.40$\\
11 &18061 & 01:00:11.942 &+28:02:25.277 & 6.33 & 26.9 & 0.08 & 1.6 & $-3.00\pm0.38$\\
12 &18268 & 01:00:07.674 &+28:03:55.250 & 6.64 & 26.7 & 0.65 & 1.7 & $-2.94\pm0.36$\\
\hline 
\label{tab:galaxies} 
\end{tabular} 
\begin{tablenotes}
  \small
  \item Columns show the galaxy number and ID, coordinates ${\rm RA}$ and ${\rm DEC}$, the galaxy's redshift as determined from its \oiii-emission lines, the observed magnitude in the NIRCam/F115W filter, its transverse distance from the central quasar, as well as the continuum $\rm S/N$ per pixel and the slope of the power-law fit to its continuum emission. 
\end{tablenotes}
\end{table*}
\end{center}

\section{Detection of the transverse proximity effect}\label{sec:trans_rp}

We search for transmitted flux from the galaxies' UV continuum emission in their spectra bluewards of their \lya emission line at wavelengths close to the systemic redshift of the foreground quasar, i.e.\ at $\lambda_{\rm obs}\approx8907$~{\AA}. We exclude the spectral range close to the galaxies' systemic redshift to avoid any contamination from the galaxies' \lya emission line. At the quasar’s redshift the universe is only $\sim870$~Myr old ($\sim 6\%$ of its current age) and thus mostly opaque to \lya photons \citep{Eilers2018a, Yang2020b, Bosman2022}, which implies that any significant detection of \lya forest transmission of the galaxies' UV continuum requires the quasar's proximity effect. 

Our observations reveal the quasar's ``light echo'', and show the expected flux transmission in the \lya forest due to the transverse proximity effect. For the first time we can observe the effect both along individual galaxy sightlines (two examples shown in Fig.~\ref{fig:data}), as well as in the unweighted stacked transmission spectrum from all twelve galaxy sightlines shown in Fig.~\ref{fig:stack}, which indicates that the quasar's radiation not only ionized the IGM along our line-of-sight but also in the transverse direction, carving out large bubbles of highly ionized gas in its environment. The flux transmission spectra along all individual galaxy sightlines at the redshift of the foreground quasar are shown in Fig.~\ref{fig:bestfit2}.

\section{Modeling the extent and geometry of the ionization cone}\label{sec:model}

The detection of the transverse proximity effect along multiple galaxy sightlines enables us to model the extent and geometry of the quasar's ionization cone. Inspired by the unification model of active galactic nuclei (AGN) \citep[e.g.][]{Antonucci1993, Urry1995}, we construct a model with a biconical ionization geometry to create a 3-dimensional map of the quasar's light echo, and to determine the onset of the quasar’s UV-luminous lifetime. 

Luminous quasars ionize their surroundings with an ionizing emissivity $\Gamma_{\rm QSO}$, which depends on the quasars' ionizing photon emission rate $\dot{N}_{\rm ion}$. The total number of ionizing photons emitted by a quasar can then be written as $N_{\rm ion}=\int\dot{N}_{\rm ion} {\rm d}t$ \citep{Davies2019b}. 
We assume a luminosity-dependent bolometric correction \citep{Runnoe2012} to convert the quasars absolute magnitude at $1450$~{\AA} in the rest-frame, i.e.\ $\rm M_{1450}=-29.26$ \citep{Wu2015}, to an ionizing luminosity assuming a spectral energy distribution (SED) \citep{Lusso2015}. 
This results in a bolometric luminosity of $L_{\rm bol}\approx9.7\times10^{47}\rm erg\,s^{-1}$, which corresponds to a rate of ionizing photons of $\dot{N}_{\rm ion}\approx1.1\times10^{58}\,\rm s^{-1}$ for the central quasar, J0100+2802. The ionizing emissivity $\Gamma_{\rm QSO}$ is linearly proportional to $\dot{N}_{\rm ion}$, i.e. $\Gamma_{\rm QSO}=\frac{-\beta\,\sigma_{912}\,\dot{N}_{\rm ion}}{(3-\beta)\,4\pi\,(d^2_\parallel+d^2_\perp)}$, where $d_\parallel$ and $d_\perp$ describe the distance along the line-of-sight and perpendicular to the quasar, respectively, while $\beta=-1.5$ and $\sigma_{912}=6.3\times10^{-18}\,\rm cm^2$ denotes the power-law slope of the quasar's ionizing spectrum and the absorption cross-section of neutral hydrogen to ionizing photons. 

The total photoionization rate $\Gamma_{\rm H\,I}$ at each spatial location in the observed volume is determined via a superposition of $\Gamma_{\rm QSO}$ and the ionizing emissivity of the UV background $\Gamma_{\rm UVB}$, which is estimated to be $\Gamma_{\rm UVB}\approx3\times10^{-13}\rm\,s^{-1}$ based on recent optical depth measurements along quasar sightlines \citep{Bosman2022}, which we extrapolate to the redshift of the quasar $z_{\rm QSO}$. This measurements corresponds to a mean transmitted flux measurement of $\langle F\rangle\approx0.0043$ \citep{Bosman2022}. 
We neglect any small intrinsic scatter due to density fluctuations and calculate the effective optical depth of the \lya forest as $\tau^\alpha_{\rm eff}\approx \gamma \times \left(\Gamma_{\rm H\,I} / 2.5\times 10^{13}\rm\,s^{-1}\right)^{-\delta}$, where the parameters $\gamma\approx6$ and $\delta\approx0.55$ were derived from a fit to an ensemble of simulated \lya forest spectra from hydrodynamical simulations \citep{Davies2019a}.

Because the quasar resides in an environment with an observed galaxy overdensity \citep{Kashino2023, Eilers2024}, we briefly assess the galaxies’ influence on the ionizing photon budget. Using the canonical value for the ionizing production efficiency, $\log_{10}{\xi_{\rm ion}}=25.2$ \citep{Matthee2017, Simmonds2024}, and assuming an ionizing escape fraction of $f_{\rm esc}\approx3\%$ \citep{Mascia2024, Papovich2025, Yue2025} we sum the UV luminosity of all galaxies within the quasar environment, i.e.\ $|z_{\rm gal}-{z_{\rm QSO}}|<0.05$, and find a photon production rate of $\dot{N}_{\rm ion, gal}=6.8\times10^{-58}\rm\,s^{-1}$, which is more than five orders of magnitude lower than the ionizing photons emitted by the quasar. We thus neglect any effects due to the galaxies in the quasar's environment. 

Our model for the quasar's ionized region describes the geometry of the ionization cone with three parameters: an inclination $i$, an orientation described by a position angle $\rm PA$, as well as an opening angle $\psi$ of the obscuring medium, which corresponds to an obscured solid angle fraction as $f_{\rm obsc}=1-\left[2\sin^2((90^\circ-\psi)/2)\right]$. The effect of each model parameter on the geometry of the ionization cone and the resulting expected transmitted flux is shown in Fig.~\ref{fig:models}. Additionally, we model the UV-luminous lifetime of the quasar $t_{\rm QSO}$ assuming a constant quasar luminosity, i.e.\ a ``light-bulb'' light curve. The onset of the quasar's UV-luminous radiation started at a time $t_{\rm QSO}$ ago, launching an ionization front that is propagating outwards with the speed of light $c$. The finite emission timescale results in a parabolic cut-off of the quasar's radiation seen from our point of view, as illustrated in the right panel of Fig.~\ref{fig:models}. 

The spatial regions within the emission cone are ionized by $\Gamma_{\rm H\,I} = \Gamma_{\rm QSO} + \Gamma_{\rm UVB}$, while the regions outside of the cones are ionized only by $\Gamma_{\rm UVB}$. Thus, any galaxy sightline piercing through the ionization cone is expected to show \lya transmitted flux around the systemic redshift of the quasar, as the residual neutral hydrogen in the IGM is ionized by the intense light from the quasar, reducing the opacity to \lya photons. The regions outside of the illuminated cone are influenced only by the UV background radiation and are thus set to the mean flux transmission of $\langle F\rangle\approx 0.0043$ at this redshift \citep{Bosman2022}. 

Each of the observed galaxies in the background of the quasar pierce through the emission cone at the sky location $(x_{\rm gal}, y_{\rm gal})$ (with the central quasar located at (0,0)), at a distance from the quasar projected into the plane of the sky of $d_\perp=\sqrt{x^2_{\rm gal} + y^2_{\rm gal}}$, while each spectral pixel along the galaxy sightline has a parallel distance to the quasar of $d_\parallel=\Delta v/H(z_{\rm QSO})$, where $\Delta v$ describes the velocity shift with respect to the systemic redshift of the quasar, and $H(z_{\rm QSO})$ denotes the Hubble constant at the quasar's redshift. We calculate the intersections between the surface of the double cone for each galaxy sightline following a model developed for the analysis of protoplanetary disks \citep{Rosenfeld2013}. 

For a given set of parameters for the ionization cone we model the expected flux profile for each galaxy given their position on the sky $(x_{\rm gal}, y_{\rm gal})$ with respect to the quasar and compare the predicted flux transmission to the observed continuum-normalized galaxy spectra. All model parameters are determined via a Markov Chain Monte Carlo (MCMC) algorithm \citep{emcee} with a likelihood function $\ln\mathcal{L}=\sum_i-\frac{(f_i-f_{{\rm model}, i})^2}{2\sigma_i^2}$, where $f_i$ and $\sigma_i$ denote the continuum-normalized flux and noise vectors of each galaxy $i$ around the quasar's systemic redshift, i.e.\ between $8500~{\rm {\AA}}<\lambda_{\rm obs}<9500~{\rm {\AA}}$ (see Fig.~\ref{fig:bestfit2}). The parameter $f_{{\rm model}, i}$ describes the expected flux transmission extracted at each galaxy position on the sky given the four model parameters, i.e. $i$, $\rm PA$, $\psi$ and $t_{\rm QSO}$. We choose flat priors for the four parameters, i.e.\ $i\in[0^\circ, 90^\circ)$, ${\rm PA}\in[0^\circ, 360^\circ)$, $\log_{10}(1-f_{\rm obsc})\in[-4, 0]$, and $\log_{10}(t_{\rm QSO}/{\rm yr})\in[4, 9]$. We apply an additional prior on the orientation of the ionization cone to ensure that our line-of-sight falls within the illuminated cone, since we would otherwise not expect to see a luminous, unobscured quasar. This latter assumption results in non-flat prior distributions for the inclination and the unobscured fraction, as shown in Fig.~\ref{fig:corner}. 

Our best model for the quasar's ionization cone is depicted in Fig.~\ref{fig:bestfit} and suggests a $1\sigma$ ($2\sigma$) upper limit on the opening angle of the obscuring medium of $\psi<65^\circ$ ($\psi<71^\circ$), corresponding to an obscured fraction of $f_{\rm obsc}< 91\%$ ($f_{\rm obsc}< 94\%$), and an onset of the UV-luminous quasar light $\log_{10}(t_{\rm QSO}/{\rm yr})=5.6^{+0.1}_{-0.3}$ ago. We also obtain weak constraints on the orientation and inclination of the ionization cone, indicating that a mildly inclined cone, i.e.\ $i\approx15^\circ$, with a position angle of $240^\circ\lesssim{\rm PA}\lesssim340^\circ$ is preferred. Fig.~\ref{fig:corner} shows the full posterior distributions of all model parameters. 

Other factors, such as variability in the quasar’s luminosity or precession of its bi-cone, may also affect the shape and extent of the surrounding ionized region. However, the limited number of bright background galaxies restricts our sensitivity to all solid angles, preventing us from constraining more complex, flexible models with additional free parameters.

\section{Implications for the early growth of black holes}\label{sec:discussion}

The estimate for the quasar lifetime $t_{\rm QSO}$ is significantly shorter than the time required to grow the quasar's approximately ten billion solar mass black hole. Assuming Eddington-limited accretion with a canonical radiative efficiency of $\epsilon\approx10\%$, we would expect the black hole accretion timescale to be comparable to the Hubble time at these redshifts, i.e. $\sim10^{8-9}$~years \citep[e.g.][]{Inayoshi2019}. Our new estimate is, however, consistent with previous measurements of the quasar's lifetime leveraging the extent of its observed proximity zone in the quasar's rest-UV spectrum along our sightline \citep{Eilers2017a, Davies2020}. It also agrees well with the recent measurement of the average UV-luminous duty cycle of high-redshift quasars, i.e.\ $f_{\rm duty}\lesssim 1\%$ \citep{Eilers2024}, inferred from the quasars' clustering strength measured by the cross-correlation of quasars with their surrounding galaxies, which represents a completely independent approach to estimate the quasars' activity timescales \citep{Eilers2024, Pizzati2024}. Furthermore, the lack of significant flux transmission in the background of the quasar, as shown in Fig.~\ref{fig:stack}, suggests that there were no significant previous UV-luminous accretion episodes due to a potentially flickering quasar light curve \citep{Davies2019a}. 
If the quasar's radiation were episodic, the emission from previous active phases would continue to propagate outwards ionizing the gas along its path, thus resulting in lower opacity ``shells'' around the quasar \citep{Adelberger2004, Bosman2019}. 

Such short UV-luminous quasar lifetimes pose significant challenges to current black hole growth models, where matter from the surrounding accretion disk is accreted with a canonical radiative efficiency of $\epsilon\approx10\%$. If a very high fraction of $>99\%$ of growing black holes in the early universe were obscured by dust or dense gas, one could reconcile the required long black hole growth timescales with the significantly shorter observed UV-luminous timescales \citep[e.g.][]{Davies2019b, Satyavolu2023}, prompting many studies to search for the predicted abundance of highly obscured quasars. However, identifying obscured high-redshift quasars with SMBHs has been challenging due to the lack of sensitive wide-field mid-infrared or X-ray surveys, and thus only very few promising candidates have been discovered to date  \citep{Endsley2022, Fujimoto2022}. 
Our detection of the transverse proximity effect of the quasar along multiple galaxy sightlines indicates that at least this high-redshift quasar is not highly obscured. We can securely rule out a solid angle obscured fraction of $>99\%$ at $>5\sigma$ significance, and our best estimate is in line with the fraction of obscured quasars at lower redshifts as determined from deep multi-wavelength wide-field surveys \citep{Polletta2008}. 

Thus, our results imply that sightline-dependent obscuration effects due to the geometry of an obscuring medium around the accreting black hole cannot exclusively account for the observed short UV-luminous black hole growth timescales of high-redshift quasars. In order to explain the existence of several billion solar mass black holes at early cosmic times, it is thus likely that these black holes accrete matter at radiatively inefficient accretion rates, i.e.\ $\epsilon\ll10\%$, as suggested in slim accretion disk models for instance \citep[e.g.][]{NarayanYi1995, BlandfordBegelman1999, BegelmanVolonteri2017}. In addition, black holes could grow in initially completely enshrouded cocoons, before their radiation pressure sheds off any obscuring medium manifesting themselves as UV-luminous quasars -- a scenario that would imply the existence of a very large number of obscured quasars hosting SMBHs in the early universe. One can speculate that the recently JWST-discovered galaxies dubbed ``Little Red Dots'' host smaller black holes in an early, obscured growth phase \citep{Matthee2023b, Naidu2025}, but a large population of obscured massive black holes is yet to be detected. 

\software{numpy \citep{numpy}, scipy \citep{scipy}, matplotlib \citep{matplotlib}, astropy \citep{astropy2013, astropy2018, astropy2022}, emcee \citep{emcee}}

\begin{acknowledgements}
This work is based on observations made with the NASA/ESA/CSA James Webb Space Telescope. The data were obtained from the Mikulski Archive for Space Telescopes at the Space Telescope Science Institute, which is operated by the Association of Universities for Research in Astronomy, Inc., under NASA contract NAS 5-03127 for JWST. These observations are associated with programs $\#1243$ and $\#4713$. 

All of the data presented in this article were obtained from the Mikulski Archive for Space Telescopes (MAST) at the Space Telescope Science Institute. The specific observations analyzed can be accessed via \dataset[doi:10.17909/w7hm-qb39]{https://doi.org/10.17909/w7hm-qb39}.

J.M. is supported by the European Union (ERC, AGENTS, 101076224).
\end{acknowledgements}

\bibliography{literature}{}
\bibliographystyle{aasjournalv7}



\end{document}